\documentclass[superscriptaddress,aps,nofootinbib,preprint]{revtex4}
\usepackage{amsmath}
\usepackage{amsfonts}
\usepackage{graphicx}
\usepackage{amssymb}

\usepackage{ifthen,color}

\newboolean{commentflag}
\setboolean{commentflag}{true}
\newcommand{\comment}[1]{\ifthenelse{\boolean{commentflag}}{{\Huge
\textcolor{red}{#1}}}{}}

\newcommand{\lag}{\ensuremath{{\mathcal{L}}}}

\newcommand{\ltsim}{\lower3pt\hbox{$\, \buildrel < \over \sim \, $}}
\newcommand{\gtsim}{\lower3pt\hbox{$\, \buildrel > \over \sim \, $}}

\begin{document}
\rightline{\textbf{UG-FT-139/02}}
\rightline{\textbf{CAFPE 9/02}}
\rightline{\textbf{FISIST/14\,-2002/CFIF}}
\vspace{.3cm}

\title{ Extra Dimensions, Isosinglet Charged Leptons and Neutrino Factories.}

\author{G.C. Branco}
\email{gbranco@cfif.ist.utl.pt} \affiliation{Centro de F\'{\i}sica das
Interac\c{c}\~{o}es Fundamentais (CFIF), Departamento de F\'{\i}sica,
Instituto Superior T\'{e}cnico, Av.Rovisco Pais, 1049-001 Lisboa,
Portugal}

\author{D. Del\'{e}pine}
\email{delepine@cfif.ist.utl.pt} \affiliation{Centro de F\'{\i}sica das
Interac\c{c}\~{o}es Fundamentais (CFIF), Departamento de F\'{\i}sica,
Instituto Superior T\'{e}cnico, Av.Rovisco Pais, 1049-001 Lisboa,
Portugal}

\author{B. Nobre}
\email{bnobre@cfif.ist.utl.pt} \affiliation{Centro de F\'{\i}sica das
Interac\c{c}\~{o}es Fundamentais (CFIF), Departamento de F\'{\i}sica,
Instituto Superior T\'{e}cnico, Av.Rovisco Pais, 1049-001 Lisboa,
Portugal}

\author{J. Santiago}
\email{jsantiag@ugr.es} \affiliation{  Departamento de F\'\i sica Te\'orica y del Cosmos and Centro Andaluz de F\'\i sica de Part\'\i culas Elementales (CAFPE), Universidad de Granada, E-18071 Granada, Spain}

\begin{abstract}
Isosinglet fermions naturally arise in a variety of extensions of the
 Standard Model, in particular in models with extra dimensions. In
 this paper, we study the effect of the addition of a new isosinglet
 charged lepton to the standard spectrum, with special emphasis on
 implications for neutrino asymmetries to be measured at future
 neutrino factories. Lepton flavour violation in neutral current and lepton universality constraints are extensively discussed. We show that new physics effects in $\nu_e-\nu_{\mu}$  $CP$ asymmetries are significantly enhanced due to leptonic maximal mixings but still too small to give a signature at future neutrino factories. A signal for $CP$ asymmetries in $\nu_{\mu}-\nu_{\tau}$ channel due to new physics  could be observed at $1-3~\sigma$ if lepton flavour violating $\tau$ decays are seen in a very close future in $B$-factories like BELLE experiment.

\end{abstract}

\maketitle


\section{Introduction}
Two of the most exciting recent developments in particle physics are, on one hand, the experimental evidence for neutrino oscillations \cite{SuperKamiokandea,MACROa} pointing towards non-vanishing neutrino masses and mixings and, on the other hand, the suggestion of models with more than three spatial dimensions \cite{extra-dimensions} which have the attractive feature of addressing, within a novel framework, long standing puzzles like, for example, the gauge hierarchy problem and the elementary fermion mass spectrum.

In this paper, we analyse some of the phenomenological consequences of isosinglet charged leptons which naturally arise in some of the models with extra-dimensions, discussing in particular the interplay between the study of lepton flavour violating rare processes and new physics effects observable in $CP$ asymmetries at neutrino factories\cite{Alsharoa:2002wu}.

In models with extra-dimensions, the use of new techniques is accompanied by a very
interesting phenomenology, with a large number of new particles populating
the spectrum at energies typically of the order of the compactification
scale. Extra-dimensional models represent then a very interesting framework
for physics beyond the SM,
with the Kaluza-Klein (KK) excitations of the
bulk fields being extra gauge bosons, fermions or scalars.

Arkani-Hamed and Schmaltz~\cite{Arkani-Hamed:1999dc} have proposed in the
context of these theories the split fermion idea, which stands for the use
of the localization properties of bulk fermion zero modes to suppress
coefficients without the need of symmetries protecting the corresponding
operators. One particularly interesting application of this idea is a
natural realisation of the observed pattern of fermion masses and mixing
angles~\cite{Mirabelli:1999ks,DelAguila:2001pu}. Recently, it was
shown~in \cite{DelAguila:2001pu} that, under very general circumstances, it
is possible to decouple the mass scale of the fermion KK excitations from
the compactification scale by means of multi-localizing the fermion zero
modes. In this way it was possible to naturally realise the observed
spectrum of quark masses and mixing angles in a model with a multi-brane
scalar background, a compactification scale $M_{c}\sim 100$ TeV~\footnote{This value is consistent with the bounds on flavour changing neutral currents in these models~\cite{Carone:1999nz}.} and still
have a light KK excitation of the right-handed (RH) top quark with
observable phenomenological consequences at present or future
colliders. 

In the class of models described in~\cite{DelAguila:2001pu}, the RH
component of the heaviest fermions are naturally multi-localized. Thus they
typically have light (vector-like) KK excitations mixing mainly with them.
One extra ingredient appearing in these models is that the SM fermion gauge
couplings are modified due to mixing with their KK excitations, with
corrections which are proportional to the masses of the SM fermions.
 We will  apply the above scenario to the leptonic
sector~\cite{Mirabelli:1999ks,DelAguila:2001pu}, analysing in detail the
phenomenological consequences of the isosinglet charged leptons which
naturally arise in this class of models. The most natural situation is to
have a multi-localized RH tau lepton with a light isosinglet KK excitation
mixing mainly with it. We will see in the sequel that, for phenomenological
reasons, it is crucial that the new vector-like charged lepton mixes mainly
with the tau.

The analysis of new physics effects in the leptonic sector, is specially relevant in view of the various experimental projects designed to uncover the neutrino world, in particular the planned neutrino factories \cite{Alsharoa:2002wu} which will measure neutrino masses and mixings with great precision and eventually $CP$ violation in the leptonic sector. Furthermore, neutrino factories have the potential to uncover new physics, which can arise in a variety of extensions of the Standard Model (SM), such as models with vector-like neutrinos \cite{delAguila:2002sx}. In the class of models with vector-like charged leptons which we consider in this paper, we will find explicit realizations of the model independent results previously found \cite{Nir2001,Ota2001}.

The paper is organised as follows. In the next section, we present a
general description of models with extra isosinglet charged leptons,
with special emphasis on lepton-flavour-violating processes and
introduce the structure motivated by models with extra dimensions.
The results
of this section are model independent and therefore  relevant for
any extension of the SM with isosinglet charged leptons. In
section III, 
we discuss the phenomenological implications of isosinglet charged
leptons on neutrino oscillations and in particular on CP asymmetries
at neutrino factories.
In section IV, we present our main results while our conclusions are
summarised in section V. We end the paper with an appendix where the
KK description of theories with extra dimensions is reviewed.

\section{General description}
In this section, first we recall  some
well-known features of models with vector-like particles, in
particular the structure of the neutral and charged flavour-changing
leptonic currents. Secondly, the constraints coming from rare
lepton-flavour violating decays and lepton universality are presented. Finally, the extra-dimensional framework for isosinglet charged lepton is described.

\subsection{Charged lepton masses and mixings.}

Let us consider the 
addition of a vector-like charged lepton, singlet under
  $SU(2)_L$, to the SM spectrum. The mass Lagrangian for the charged leptons 
can be written with complete generality in
  the current eigenstate basis as

\begin{equation}
\mathcal{L}_{m}=\left( 
\begin{array}{ll}
\overline{l}^\prime_{L} & \overline{L}_{L}
\end{array}
\right) \left( 
\begin{array}{ll}
m_{l} & J \\ 
0 & M
\end{array}
\right) \left( 
\begin{array}{l}
l_{R} \\ 
L_{R}
\end{array}
\right) +h.c.,  \label{charged lepton mass matrix 1}
\end{equation}
where $m_{l}$ is a $3\times 3$ mass matrix and $J$ describes the mixing between
the first 3 families and the vector-like charged
lepton. $L_{L,R}$ 
are the Left-Handed (LH) and
Right-Handed (RH) components of the isosinglet
charged lepton and 
$l^\prime_{L},l_R$ the usual LH and RH charged lepton
fields respectively ($e,\mu ,\tau $).
Note that the $(1 \times 3)$ zero matrix in eq.(\ref{charged lepton mass matrix 1}) corresponds to an allowed choice of weak-basis and does not imply any loss of generality. 
The matrices  $m_l$ and $J$  are $\Delta I=1/2$ 
mass terms, therefore proportional to the SM Higgs vacuum
expectation value while $M$ is an
$SU(2)_{L}\otimes U(1)_{Y}$ invariant mass terms, which can be arbitrarily
large, since it is not protected by the low energy gauge
symmetry.

Before diagonalising the charged lepton mass matrix, let us have a look at
the neutrino mass sector. We shall assume that naturally small LH 
neutrino Majorana masses are generated through the breaking of lepton number
at high energy. It is well known that the seesaw mechanism provides
one of the most attractive scenarios for generating small LH 
neutrino masses \cite{seesaw}. However, for our discussion, the detailed
origin of neutrino masses is not important. The Majorana mass term for the
3 light LH 
neutrinos can be written as 
\begin{equation}
\mathcal{L}_{\nu }^{\Delta L=2}=\frac{1}{2}\overline{\nu }_{L}^{c}M_{\nu
}\nu _{L}+h.c.,
\end{equation}
with $M_{\nu }$ 
a $3\times 3$ symmetric matrix which can be diagonalised
by a unitary transformation $U_{\nu }$
\begin{equation}
U_{\nu }^{T}M_{\nu }U_{\nu }\equiv diag(m_{\nu _{e}},m_{\nu _{\mu }},m_{\nu
_{\tau }}).
\end{equation}
We can go to the mass eigenstate basis for the LH neutrinos by making
the corresponding unitary transformation on the lepton doublets
\begin{equation}
\left( 
\begin{array}{l}
\nu _{L}^{m} \\ 
l_{L}
\end{array}
\right) =U_{\nu }^{\dagger }\left( 
\begin{array}{l}
\nu _{L} \\ 
l_{L}^{^{\prime }}
\end{array}
\right).
\end{equation}
Of course, eq.(\ref{charged lepton mass matrix 1}) is not invariant under
this transformation. In the neutrino
mass eigenstate 
basis, the mass matrix
for the charged leptons is given now by 
\begin{equation}
\mathcal{L}_{m}=\left( 
\begin{array}{ll}
\overline{l}_{L} & \overline{L}_{L}
\end{array}
\right) \left( 
\begin{array}{ll}
U_{\nu }^{\dagger }m_{l} & U_{\nu }^{\dagger }J \\ 
0 & M
\end{array}
\right) \left( 
\begin{array}{l}
l_{R} \\ 
L_{R}
\end{array}
\right) +h.c.\equiv (\overline{l}_{L})M_{l}(l_{R}).  \label{mass lepton2}
\end{equation}
It may seem not necessary to discuss the neutrino sector at this point.
Indeed, without loss of generality, 
we could have
 started 
our discussion with eq. (%
\ref{charged lepton mass matrix 1}) in the physical basis for neutrinos.
However, models where vector-like fermions naturally appear (e.g. models
with extra-dimensions) often predict some textures for $J$, $m_{l}$ or
$M_{\nu }$. In such cases, it is convenient to work with eq.(\ref{mass lepton2}%
) where 
the constraints on the structure of $J,m_{l}$ and $M_\nu$ can be
 implemented in a straigthforward way.

The charged lepton mass matrix
$M_{l}$ is diagonalized by the
unitary transformations $U_{L}$ and $%
U_{R}$
\begin{equation}
U_{L}^{\dagger }M_{l}U_{R}=\mathrm{diag}(m_{e},m_{\mu },m_{\tau },M_{D}),
\end{equation}
with the mass eigenstates $l_{L,R}^{m}$ given by 
\begin{equation}
(l_{L,R})=U_{L,R}(l_{L,R}^{m}).
\end{equation}
The left rotation
$U_{L}$ can be written as~\cite{branco1986,branco1999} 
\begin{equation}
U_L=\left( 
\begin{array}{cc}
U_\nu^\dagger K & U_\nu^\dagger J/M \\
-J^\dagger K/M & 1
\end{array} \right)
+O\left(\frac{m_l}{M}\right)^2, \label{UL:approx}
\end{equation}
where $K$ is  
the unitary matrix which
diagonalizes $m_{l}m_{l}^{\dagger }$.

\subsection{Z-mediated flavour changing neutral current interactions.}

The leptonic neutral current gauge interaction 
reads, in the weak eigenstate basis,
\begin{equation}
\mathcal{L}_{Z}=\frac{g}{\cos\theta _{w}}Z_{\mu }(J_3^\mu -\sin^{2}
\theta _{w}J_{em}^{\mu}),
\end{equation}
with 
\begin{eqnarray}
J_{3}^{\mu }&=&\frac{1}{2}\overline{\nu }_{L}\gamma ^{\mu }\nu
_{L}-\frac{1}{2} \overline{l}_{L}\gamma ^{\mu }l_{L},
\\
J_{em}^{\mu }&=&-(\overline{l}\gamma ^{\mu }l+\overline{L}\gamma ^{\mu }L).
\end{eqnarray}
In the mass eigenstate basis, for the
  light charged leptons, one gets
\begin{equation}
\mathcal{L}_Z^\mathit{light}=
-
\frac{g}{2\cos\theta _{w}}
Z_\mu \overline{l}_{Li}^{m}[\delta_{ik} (1-2 \sin^2 \theta_w) 
-
\beta_{ik} ] \gamma^\mu 
l_{Lk}^{m},
 \end{equation}
with $i,k=1,2,3$, $\beta_{ik} \equiv U_{L4i}^{*}U_{L4k}$. 
The effect of the vector-like singlet is
  the appearance of 
the flavour changing neutral couplings (FCNC) $\beta_{ik}$ at tree level.
Note that at first order in $m_l^2/M^2$, 
\begin{eqnarray}
\beta _{ij} 
&= &
\frac{J_{l}K_{li}^{*}J_{k}^{*}K_{kj}}{M^{2}},
\end{eqnarray}
and thus these couplings are  naturally suppressed if $J_l\ll M$.

\subsection{ Higgs mediated FCNC.}

The interaction between charged leptons and the neutral Higgs boson is
given, in the weak eigenstate basis, by
\begin{equation}
\mathcal{L}_{H}=-\frac{g}{2M_{W}}\left[ \overline{l}_{L}m_{l}l_{R}+\overline{%
l}_{L}JL_{R}+h.c.\right] H^{0}.
\end{equation}
After diagonalisation of the mass matrix, 
 the Lagrangian for the light charged leptons becomes
\begin{equation}
\mathcal{L}_{H}^\mathit{light}=
-\frac{g}{2M_{W}}\left[ \overline{l}^{m}_{Li} \left(m_i \delta_{ik}-M U_{L4i}^* U_{R4k} \right) l^m_{Rk}+h.c.\right] H^{0}, 
 \label{higgscoupling}
\end{equation}
where $i,k=1,2,3$ and $m_{1,2,3}\equiv m_{e,\mu,\tau}$.
The interaction with the pseudoscalar neutral field $\chi $ is given
by 
\begin{equation}
\mathcal{L}\chi =-\frac{ig}{2M_{W}}\left[ \overline{l}_{L}m_{l}l_{R}+%
\overline{L}_{L}Jl_{R}-h.c.\right] \chi.
\end{equation}
Proceeding in the same way as for the neutral Higgs scalar, the
interaction with the pseudoscalar Higgs field $\chi$ is given by 
\begin{equation}
\mathcal{L}_{\chi}^\mathit{light}=
-\frac{ig}{2M_{W}}\left[ \overline{l}^{m}_{Li} \left(m_i \delta_{ik}-M U_{L4i}^* U_{R4k} \right) l^m_{Rk}-h.c.\right] \chi.  
\end{equation}

\subsection{Flavour changing charged current interactions.}

The interaction with the $W^{\pm }$ is given as usual by 
\begin{equation}
\mathcal{L}_{W}=\frac{g}{\sqrt{2}}(W_{\mu }^{-}J^{\mu +}+h.c.),
\end{equation}
with 
\begin{equation}
J^{\mu +}=\overline{l}_{L}\gamma^{\mu }\nu_{L}.
\end{equation}
In the mass eigenstate basis, the charged current interaction reads
\begin{equation}
\mathcal{L}_{W}=\frac{g}{\sqrt{2}}(W_{\mu }^{-}\overline{l}_{L\alpha
}^{m}U_{L\alpha k}^{\dagger }\gamma ^{\mu }\nu _{Lk}^{m}+h.c.), \label{cc1}
\end{equation}
with $\alpha =1\ldots 4$ and $k=1
\ldots 3$.
The $V_{MNS}$ mixing matrix is therefore $4\times 3$ and given by 
\begin{equation}
(V_{MNS})_{\alpha k}\equiv \left( U_{L}^{\dagger }\right) _{\alpha k}.
\end{equation}

\subsection{Limits from rare lepton flavour changing decays and lepton universality.}

Lepton-flavour violating processes are strongly constrained by the
experimental limits on rare $\mu $ and $\tau $ decays (see for
instante Table~\ref{experimental:bounds:table}).
In the present model, Z-mediated
FCNC induce tree level corrections to processes 
like $\mu \rightarrow 3e$ or $\mu
\rightarrow e$ in $_{22}^{48}$Ti while
$\mu \rightarrow e\gamma $ receives corrections at one-loop level. 
We shall consider first the limits coming from tree-level rare tau and muon
flavour changing decays and from gauge coupling lepton universality. 
Limits coming from loop-induced processes like $\mu
\rightarrow e\gamma $ will be discussed afterwards.

\begin{table}[ht]
\caption{Experimental limits on rare $\mu$ and $\tau$ decays constraining 
  FCNC.\label{experimental:bounds:table}} 
\begin{center}
\begin{tabular}{|cc|}\hline
$Br(\mu \rightarrow e\gamma )<1.2\times 10^{-11}$ & \cite{mega} \\ 
$Br(\mu \rightarrow 3e)<1.0\times 10^{-12}$ & \cite{sindrum} \\ 
$Br(\mu \rightarrow e\mbox{ in }_{22}^{48}\mbox{Ti})<6.1\times 10^{-13}$ & %
\cite{sindrum2} \\ 
$Br(\tau \rightarrow 3e)<7.8\times 10^{-7}$ & \cite{BELLE} \\ 
$Br(\tau \rightarrow 3\mu )<8.7\times 10^{-7}$ & \cite{BELLE}
\\ \hline
\end{tabular}
\end{center}
\end{table}

\subsubsection{ $\tau $ and $\mu $ rare flavour changing decays and lepton
universality}

The restrictions on rare lepton decays can be translated to indirect
bounds on lepton flavour violating branching ratios for $Z$ decays
\cite{vissani2001},
which are directly connected to the parameters of the Lagrangian. For
instance the branching ratio for $Z\to l_i l_j$ is
approximately given, for $i\neq j$, by

\begin{equation}
Br(Z\rightarrow l_i l_j)\simeq \frac{1}{8}\left| \beta_{ij}\right|
^{2}.
\end{equation}
Using the current experimental limits, 
one obtains
\begin{equation}
\begin{array}{ll}
\left| \beta _{\mu e}
\right| \leq 1.1\times 10^{-6} 
&
\mbox{from }
(\mu \rightarrow 3e),
\\ 
\left| \beta _{\mu e}\right| \leq 4.0\times 10^{-7} 
&
\mbox{from }
(\mu \rightarrow e\mbox{ in }_{22}^{48}\mbox{Ti}),
\\ 
\left| \beta
_{\tau e}\right| \leq 2.3\times 10^{-3} 
&
\mbox{from }
(\tau \rightarrow 3e),
\\ 
\left| \beta
_{\tau \mu }\right| \leq 2.4\times 10^{-3}
&
\mbox{from }
(\tau \rightarrow 3\mu ).
\end{array}
\label{bound1}
\end{equation}

Lepton universality sets limits on the diagonal elements
$\beta_{ii}$. Indirect bounds from unitarity loss in charged currents
have been computed for the case of vector-like
neutrinos~\cite{delAguila:2002sx} but give much less stringent
constraints than direct violations of universality appearing in the
model with extra isosinglet charged leptons. 
Using the one sigma
deviations for the effective charged lepton couplings,
$g_{V,A}^{e,\mu,\tau}$, to the
$Z$~\cite{LEPEWWG:web} 
as an estimation of the allowed new physics
contribution we get
\begin{eqnarray}
\left| \beta _{ee}\right| &\leq &0.0007,  \nonumber \\
\left| \beta _{\mu \mu }\right| &\leq &0.0011,  \label{universality} \\
\left| \beta _{\tau \tau }\right| &\leq &0.0013.  \nonumber
\end{eqnarray}
The last two limits are restrictive enough to impose a bound on the
$\beta_{\mu \tau}$ coefficient stronger than the one obtained from
rare tau decays
\begin{equation}
\left| \beta_{\mu \tau} \right|
=\sqrt{
\left| \beta_{\mu \mu} \right|
\left| \beta_{\tau \tau} \right|
}
\leq
1.2 \times 10^{-3}.
\end{equation}

These limits can be translated to 
constraints on the elements of the charged lepton mass matrix making
further assumptions.
As examples, we shall consider two theoretically motivated
ans\"atze 
for $J_{i}$.

\medskip

\noindent \textbf{Case A}

Given the fact that experimental constraints are much more stringent for the
first families than for the heavier ones and that both the entries in the
mass matrix $J$ and the standard lepton masses arise from Yukawa
couplings, a reasonable ansatz for the values of the $J_{l}$ is~\cite
{delAguila:fs} 
\begin{equation}
\frac{J_{i}}{J_{l}}\sim \frac{m_{i}}{m_{l}}.  \label{mass ratio fo J}
\end{equation}
Using eq.(\ref{mass ratio fo J}) and assuming that the mixing induced by $K$
is negligible, it is possible to extract limits on $J_{i}/M$ from the
experimental bounds given in eqs.(\ref{bound1},\ref{universality}),
the strongest bounds being
 \begin{eqnarray}
S_1 \equiv \frac{J_{1}}{M}&\leq& 4.4\times 10^{-5},
\nonumber
\\
S_2 \equiv \frac{J_{2}}{M}&\leq& 0.91\times 10^{-2},
\label{bound2}
\\
S_3 \equiv \frac{J_{3}}{M}&\leq& 0.036.  \nonumber 
\end{eqnarray}
These constraints come from $\mu \rightarrow e$ conversion in nuclei
for the first two families and from 
lepton universality 
for the third one.

\medskip

\noindent\textbf{Case B}

It is however possible to use a different ansatz, motivated by models with
extra dimensions~\cite{DelAguila:2001pu,delAguila:2000kb} reviewed in
the Appendix, which leads to a natural suppression for the corrections
to the couplings of the first family.
Let us assume that in the
weak basis, the vectors $J_{i}$ and $(m_{l})_{i3}$ are parallel, 
\begin{equation}
J_{i}=\lambda (m_{l})_{i3},\label{extra-dim:ansazt}
\end{equation}
with $\lambda $ a flavour independent constant.
In this situation we can use the following property to compute the value for
the $\beta _{ij}$ parameters, 
\begin{equation}
(J^{\dagger }K)_{i}^{* }=K_{ij}^{\dagger }J_{j}=\lambda K_{ij}^{\dagger
}(m_{l})_{j3}=\lambda \,m_{i}(K^{R})_{i3}^{\dagger
  }+O\left(\frac{m_l}{M}\right)^2,
\end{equation}
with no summation on the index $i$ and where $m_{i}$ are the mass
eigenvalues of the charged leptons and $K^{R}$ is defined by 
\begin{equation}
K^{\dagger }m_{l}K^{R}\equiv diag(m_{e},m_{\mu },m_{\tau })+
O\left(\frac{m_l}{M}\right)^2.
\end{equation}
Then, we can write 
\begin{equation}
\beta _{ij}=\lambda ^{2}\frac{m_{i}(K^{R})_{i3}^{\dagger }K_{3j}^{R}m_{j}}{%
M^{2}}.  \label{betaij1}
\end{equation}
The ratio of FCNC processes is then not only suppressed by masses but also
by mixing angles. For instance, 
\begin{equation}
\frac{|\beta _{e\mu }|}{|\beta _{\mu \tau }|}=\frac{m_{e}|K_{3e}^{R}|}{%
m_{\tau }|K_{3\tau }^{R}|},  \label{light:suppression}
\end{equation}
and, apart from the mass suppression, there is an extra suppression 
by mixing angles if $|K_{3e}^{R}|\ll |K_{3\tau }^{R}|$.


\subsubsection{$\mu \rightarrow e\gamma $}

This decay is induced at one-loop level,  
the constraints coming
from this process are thus expected 
to be less stringent than the ones  
arising from the
processes studied 
in the previous section. Nevertheless, the 
experimental progress on $\mu
\rightarrow e\gamma $ 
expected in the next few years \cite{PSI}, makes it worth
  studying it in detail. 
The branching ratio for
the transition $\mu \rightarrow e\gamma $ has the form 
\begin{equation}
Br(\mu \rightarrow e\gamma )=384\pi ^{3}\alpha \frac{v^{4}}{m_{\mu }^{2}}%
\left| A(m_{\mu }^{2})\right| ^{2},
\end{equation}
where $v=(8G_{F}^{2})^{-1/4}\simeq 174$ GeV is the Higgs vacuum expectation
value, $\alpha$ is the fine structure constant and
$A(q^{2})$ the form factor coming from the one-loop computation. 
At one-loop order, the transition amplitude is induced by 3 differents
classes of Feynmann diagrams respectively due to $W$, $Z$ and Higgs
exchange. For diagrams with $W$ exchange, the photon line is attached to the 
$W$'s. This diagram is suppressed in the SM by neutrino
masses whereas in our model, due to the non-unitarity of the $V_{MNS}$, it
gives an independent contribution. Let us consider the form
factors arising from these three kinds of diagrams\footnote{%
See ref.\cite{chang2000} for details on the form factor computation.
}
\begin{equation}
A=A_{W}+A_{Z}+A_{H}.
\end{equation}
We evaluate these form factors in the limit $m_{l,\nu
}^{2}/m_{W,Z,H}^{2}\rightarrow 0$ for the light families
and we assume that in our model  $M_{D} \gg m_{W,Z,H}$ for the isosinglet charged lepton contribution \footnote{ Indeed, experimental search on heavy charged lepton imposes $M_D >100.8$ GeV \cite{PDG}.}. 
They read, in this limit,

\begin{eqnarray}
A_{W} &\simeq &\frac{G_{F}}{2\pi ^{2}\sqrt{2}}m_{\mu }U_{L4\mu
}U_{L4e}^{*}\left( -\frac{5}{12}\right),  \\
A_{Z} &\simeq &\frac{G_{F}}{2\pi ^{2}\sqrt{2}}m_{\mu }U_{L4\mu
}U_{L4e}^{*}\left( \frac{3}{8}-\frac{13}{48}\frac{m_{Z}^{2}}{M_{D}^{2}}%
\right),  \\
A_{H} &\simeq &\frac{G_{F}}{2\pi ^{2}\sqrt{2}}m_{\mu }U_{L4\mu
}U_{L4e}^{*}\left(- \frac{1}{6}\frac{m_{H}^{2}}{M_{D}^{2}}\right). 
\end{eqnarray}
The branching ratio for $\mu \rightarrow e\gamma $ is then given by 
\begin{equation}
Br(\mu \rightarrow e\gamma )\simeq \frac{6\alpha }{\pi }\left| U_{L4\mu
}U_{L4e}^{*}\right| ^{2}\left( \frac{1}{24}+\frac{13}{48}\frac{m_{Z}^{2}}{%
M_{D}^{2}}+\frac{1}{6}\frac{m_{H}^{2}}{M_{D}^{2}}\right)^{2},
\end{equation} 
Using the strongest
limit on $\left| U_{L4\mu }U_{L4e}^{*}\right| $ implied by eqs.(\ref{bound1}%
) and the fact that $M_{D}^{2}/m_{Z,H}^{2}$ $\gtrsim 1$, 
this branching ratio 
is bounded by
\begin{equation}
Br(\mu \rightarrow e\gamma )\leq 5.5\times 10^{-16},
\end{equation}
far below the sensitivity of current and planned experiments. As expected, 
$\mu \to e \gamma$ does
not introduce any significant restriction, in a  model with  isosinglet charged leptons.

\subsection{Extra-dimensional framework for isosinglet charged leptons}

As we have emphasized in the introduction, theories with extra
dimensions provide an exciting arena for the study of physics beyond
the SM. It has been recently shown that models with multi-brane
backgrounds~\cite{DelAguila:2001pu} naturally present 
light isosinglet fermions which correspond to the first KK
excitations of the heavier SM weak singlets ($\tau_R$ in the leptonic
sector). We review in the appendix the details of the KK description
of the theory, the relevant features at this point being the low
energy spectrum and the form of the Yukawa couplings. In the case of
interest, the spectrum consists on the fermion zero modes, which are
the SM fermions, plus one relatively light (as compared to the
compactification scale) isosinglet charged lepton, mixing mainly with
the tau. The rest of the spectrum have masses of the order or the
compactification scale which decouples from the low energy physics.
The other feature relevant for our discussion is the form of the
Yukawa couplings. We take the scalar field responsible for the
electroweak symmetry breaking to live in a four-dimensional boundary. The
effective Yukawa coupling for the $n-$th and $m-$th modes 
satisfy the following factorisation property 
\begin{equation}
\lambda^{e(nm)}_{ij}=
\frac{\lambda^{(5)e}_{ij}}{\pi R} f^{l_i(n)}_L(0) f^{e_j(m)}_R(0),
\end{equation}
where 
$\lambda^{(5)e}_{ij}$ is the five-dimensional Yukawa
coupling and, for definiteness, we have taken the Higgs field to live in the $%
y=0 $ boundary. 
Notice that this factorisation property implies the relation in
  eq.(\ref{extra-dim:ansazt}), with
\begin{equation}
\lambda=\frac{f^{e_3(1)}_R(0)}{f^{e_3(0)}_R(0)}.
\end{equation} 
As we have seen above, this property has a
very relevant feature from the phenomenological point of view, which is the
fact that corrections on the fermion zero mode gauge couplings due to the
mixing with the heavy vector-like excitations scale with the masses of the
fermion zero modes and are thus only relevant for the heavy fermions~\cite
{delAguila:2000kb}.

It is now clear how the hierarchical pattern of charged lepton masses is
generated. Starting with five-dimensional (thus dimensionful) Yukawa
couplings of natural order, $\lambda^{(5)e}_{ij}\sim \pi R$, we can generate
effective Yukawa couplings of the appropriate size modifying the
localization properties of the different fields and thus their overlapping
with the Higgs boundary. Given the relatively large mass of the tau lepton
it is natural to have its RH component strongly localized near the Higgs
boundary and thus multi-localized (see the Appendix) 
in the set-up we are considering~\footnote{%
Notice that from the tiny neutrino masses we could expect the third family
doublet not to be too strongly localized at the Higgs boundary, thus the
mass of the tau being essentially generated by the order one overlapping of
its RH component.}.

If we ask for the RH tau lepton to be strongly multi-localized, so that it
will have a light KK excitation, then it is natural to have all mixing
angles in the charged leptonic sector small but the one in the $(2,3)$ LH
sector, which can vary, depending on the specific values of the parameters,
from moderately small to maximal. In the first case all bi-maximal mixing
has to be generated from the neutrino sector (what can be easily
accomplished in the model we have~\cite{Santiago:thesis}) whereas in the
second one only the $(1,2)$ large mixing comes from the neutrino
sector.

\section{Effects of vector-like charged leptons on $\nu $ oscillations}

\subsection{ General parametrisation}

The new interactions generated by the extra vector-like charged lepton
have important effects on neutrino oscillations. Similar effects occur
in the charged current sector in models with extra isosinglet
neutrinos~\cite{delAguila:2002sx,Langacker:1988up} while neutral
current processes could allow to discriminate between them. 
In this section we compute the effects of the
vector-like charged leptons on neutrino oscillations.

As we have seen in the previous sections, in the case of one vector-like
charged lepton, the leptonic mixing in charged currents
is described by a $4 \times
3 $ matrix $V_{MNS}$. 
The mixing among the light families is given by a
non-unitary $3\times 3$ submatrix of $V_{MNS}$ which,
using that deviations from unitarity are necessarily small, 
we will separate in two terms 
\begin{equation}
(V_{MNS})_{ik}=(U_{MNS})_{ik}+(\delta U)_{ik}, \quad i,k=1,2,3,  \label{UMNS}
\end{equation}
where $U_{MNS}$ is a $3 \times 3$ 
unitary matrix which accounts, at leading order, for
the mixing among the light families \cite{Maki:mu} and $\delta U$ describes the small
deviations from unitarity. 
In the following we consider the mass
eigenstate basis for the charged leptons and the flavour basis for
neutrinos, defined as
\begin{equation}
\nu _{L}^{f}\equiv U_{MNS}\nu _{L}^{m}.  \label{flavor1}
\end{equation}
The charged current interaction reads in this basis, for the light
leptons, 
\begin{eqnarray}
\mathcal{L}_{W}^\mathit{light} 
=\frac{g}{\sqrt{2}}\left[ W_{\mu }^{-}\overline{l}_{Li}^{m}
\left( \delta_{ik}+A^{NP}_{ik} \right)
\gamma ^{\mu }\nu _{Lk}^{f}+h.c.\right]
\equiv \mathcal{L}_{W}^{SM}+\mathcal{L}_{W}^{NP}, 
\label{SM} 
\end{eqnarray}
where $\mathcal{L}_{W}^{SM}$, the term proportional to the identity, 
is equivalent to the SM charge current interaction
and $\mathcal{L}_{W}^{NP}$ describes the New Physics effects due to the
isosinglet charged lepton with coefficient $A^{NP}_{ik}=\delta U_{ij}
(U^\dagger_{MNS})_{jk}$.

In order to compute the effect of New Physics, we need to know $U_{MNS}$ and 
$\delta U$. For that, we have to diagonalise the charged lepton mass
matrix given in eq.(\ref{mass lepton2}). To define the SM families,we
shall proceed in 2 steps: first, $M_{l}$ is diagonalised in blocks so
that the
heavy charged lepton mass eigenstate is defined. In the second step 
we make a unitary transformation involving only the light charged leptons,
which completes the diagonalization. This stepwise diagonalisation
corresponds to making the following sequence of rotations, 
for the LH sector,
\begin{equation}
U_{L}\equiv R_{34}\widetilde{R}_{24}\widetilde{R}_{14}R_{12}^{\dagger }%
\widetilde{R}_{13}^{\dagger }R_{23}^{\dagger },  \label{UL1}
\end{equation}
where 
\begin{equation}
\widetilde{R}_{13}=\left( 
\begin{array}{llll}
c_{13} & 0 & s_{13}e^{-i\delta _{13}} & 0 \\ 
0 & 1 & 0 & 0 \\ 
-s_{13}e^{i\delta _{13}} & 0 & c_{13} & 0 \\ 
0 & 0 & 0 & 1
\end{array}
\right) , \quad R_{ij}\equiv \widetilde{R}_{ij}(\delta _{ij}=0).
\end{equation}
An
explicit form for $U_{L}$ can be found in ref.\cite{Fritzsch1987}. It is
easy to check that making the rotation in the order defined in eq.(\ref{UL1}%
), we have 
\begin{eqnarray}
U_{L}^{\dagger }M_{l}M_{l}^{\dagger }U_{L} &=&R_{23}\widetilde{R}_{13}R_{12}%
\widetilde{R}_{14}^{\dagger }\widetilde{R}_{24}^{\dagger }R_{34}^{\dagger
}M_{l}M_{l}^{\dagger }R_{34}\widetilde{R}_{24}\widetilde{R}%
_{14}R_{12}^{\dagger }\widetilde{R}_{13}^{\dagger }R_{23}^{\dagger }
\nonumber \\
&=&R_{23}\widetilde{R}_{13}R_{12}\left( 
\begin{array}{ll}
\overline{m}_{l}^{2} & 0 \\ 
0 & M_{D}^{2}
\end{array}
\right) R_{12}^{\dagger }\widetilde{R}_{13}^{\dagger }R_{23}^{\dagger }.
\end{eqnarray}
Thus $R_{23}\widetilde{R}_{13}R_{12}$ is the mixing matrix connecting the 3  
light families. It is important to recall that since $M_{D}\gg
J\gg m_{\tau ,\mu ,e}$, the mixing angles $\theta _{i4}$ ($i=1,2,3$) are
well approximated by 
\begin{equation}
\theta _{i4} e^{i\delta_{i4}}\simeq \frac{(U_{\nu }^{\dagger }J)_{i}}{M}. 
\label{thetai4_def}
\end{equation}
With this prescription,
the $U_{MNS}$ matrix as defined by eq.(\ref{UMNS}) is given by 
\begin{equation}
(U_{MNS})_{ik}=(R_{23}\widetilde{R}_{13}R_{12})_{ik},
\end{equation}
and the New Physics contribution by
\begin{equation}
A^{NP}_{ik}=\delta U_{ij}(U^\dagger_{MNS})_{jk}=
\left( U_{MNS}
\widetilde{R}_{14}^{\dagger }\widetilde{R}_{24}^{\dagger }R_{34}^{\dagger
} U_{MNS}^\dagger \right)_{ik}-\delta_{ik}.
\end{equation}
Using the above parametrisation, $U_{MNS}$ can be expressed as 
\begin{equation}
U_{MNS}=\left( 
\begin{array}{lll}
c_{12}c_{13} & c_{13}s_{12} & s_{13}e^{-i\delta _{13}} \\ 
\begin{array}{l}
-c_{23}s_{12}-c_{12}s_{13}s_{23}e^{i\delta _{13}}
\end{array}
& 
\begin{array}{l}
c_{12}c_{23}-s_{12}s_{13}s_{23}e^{i\delta _{13}}
\end{array}
& c_{13}s_{23} \\ 
\begin{array}{l}
-c_{12}c_{23}s_{13}e^{i\delta _{13}}+s_{12}s_{23}
\end{array}
& 
\begin{array}{l}
-c_{12}s_{23}-c_{23}s_{12}s_{13}e^{i\delta _{13}}
\end{array}
& 
\begin{array}{l}
c_{13}c_{23}
\end{array}
\end{array}
\right) P,
\end{equation}
where $P=\mbox{diag}(1,e^{i\beta _{1}},e^{i\beta _{2}})$. The phases $\beta
_{1,2}$ are Majorana-type phases which will not play any r\^{o}le in our
discussion.
The New Physics contribution can be also explicitly written as
\begin{equation}
A^{NP}=U_{MNS}\left( 
\begin{array}{lll}
c_{14} & -s_{14}s_{24}e^{-i(\delta _{14}-\delta _{24})} & 
-s_{14}s_{34}c_{24}e^{-i\delta _{14}} \\ 
0 & c_{24} & -s_{24}s_{34}e^{-i\delta _{24}} \\ 
0 & 0 & c_{34}
\end{array}
\right) U_{MNS}^{\dagger }-I,  \label{ANPbis}
\end{equation}
with $I$ denoting the $3\times 3$ identity matrix.

Note that the full $4\times 3$ $V_{MNS}$ matrix contains, apart from $%
\delta _{13}$, two extra Dirac-type $CP$ violating phases. These phases
appear always in combination with the small mixing angles $\theta _{i4}$. The
appearance of these two extra Dirac-type phases has to do with the fact that
the $3\times 1$ column matrix $J$ is complex, with one of the phases
eliminated through the rephasing of the isosinglet charged lepton field \cite
{branco1986}.

\medskip 
Once we have separated the New Physics contributions from the SM
Lagrangian, we can use the formalism described in
\cite{Nir2001,Ota2001} to compute the effect 
of this new interaction on
neutrino asymmetries.

In neutrino factories,  neutrinos are usually produced through muon
decay. Therefore, assuming that we have a $\mu ^{+}$ beam, 
typically the production process of a neutrino state, $\nu
_{\alpha }$, in conjunction with a $e^{+}$ is given by 
\begin{equation}
\mu ^{+}\rightarrow e^{+}\nu _{\alpha }\overline{\nu }_{\mu }.
\label{nuprocess}
\end{equation}
The production of a neutrino state $\nu
_{e}$ in conjunction with a $e^{+}$ is also possible
through the process
\begin{equation}
\mu ^{+}\rightarrow e^{+}\nu _{e}\overline{\nu }_{\alpha }. \label{nualpha}
\end{equation}
The detector process, for a wrong sign event, 
is $\nu _{\rho }d\rightarrow \mu ^{-}u$. Between
production and detection, the oscillations of $\nu _{\alpha }$ from (\ref
{nuprocess}) or of $\nu _{e}$ from (\ref{nualpha}) into a $\nu _{\rho }$ can
take place.
The important point to note is that only the process given in (\ref
{nuprocess}) is able to interfere with the SM process, $\mu ^{+}\rightarrow
e^{+}\nu _{e}\overline{\nu }_{\mu }$, due to the fact that they have the
same initial and final states $\mu ^{+}$, $e^{+}$ and $\overline{\nu }_{\mu
} $. So in the sequel, we shall neglect the contribution coming from process
(\ref{nualpha}) as it does not interfere with SM amplitude,
\textit{i.e.} its effects are
higher order in New Physics compared to process (\ref{nuprocess}).

This approach can be generalised
 to any flavour and to any production and detection processes \cite{Ota2001}.
Consider
the Lagrangian responsible for the production of a $\nu _{\beta }$ in
conjunction with a $l_{\alpha }^{+}$, 
\begin{equation}
\mathcal{L}^{s}=2\sqrt{2}G_{F}
(\delta_{\alpha \beta}+\epsilon _{\alpha \beta }^{s})
\left( \overline{%
\mu }\gamma ^{\mu }P_{L}\nu _{\mu }\right) \left( \overline{\nu}_{\beta}%
\gamma _{\mu }P_{L}l_{\alpha }\right), 
\end{equation}
where $P_{L}$ is defined as usual by $(1-\gamma _{5})/2$. According
our notation, 
\begin{equation}
\epsilon^{s}_{\alpha \beta}=
(A^{NP*})_{\alpha \beta}.
\end{equation}
For the detection of a $\nu _{\beta }$ signalled in the detector by a $%
l_{\alpha }^{-}$, the similar 4 Fermi interaction enters 
\begin{equation}
\mathcal{L}^{d}=2\sqrt{2}G_{F}
(\delta_{\alpha \beta}+\epsilon _{\alpha \beta }^{d})
\left( \overline{u%
}\gamma ^{\mu }P_{L}d\right) \left( \overline{l}_{\alpha }\gamma _{\mu
}P_{L}\nu _{\beta }\right).
\end{equation}
with 
\begin{equation}
\epsilon^{d}_{\alpha \beta}=A_{\alpha \beta
}^{NP}\Rightarrow \epsilon^{d}=\epsilon^{s* }.
\end{equation}

As done in ref.\cite{Nir2001},  
we define $\nu _{e}^{s}$ as the neutrino
state that is produced in the source in conjunction with a $e^{+}$,
the production process being $\mu ^{+}\rightarrow e^{+}\nu _{\alpha
}\overline{\nu }_{\mu }$, and $\nu _{\mu }^{d}$ the neutrino state 
that is signalled by $\mu ^{-}$ production in the detector, the detector process being $\nu _{\alpha}d\rightarrow \mu ^{-}u$. The 
detector process for a wrong sign event 
is interpreted as due
to the oscillation of $\nu _{e}$  into $\nu
_{\mu }$.

\begin{eqnarray}
\left| \nu _{e}^{s}\right\rangle  &=&\sum_{i}\left( (1+\epsilon
_{ee}^{s})(U_{MNS})_{ei}^{*}
+\epsilon _{e\mu }^{s}(U_{MNS})_{\mu i}^{*}+\epsilon
_{e\tau }^{s}(U_{MNS})_{\tau i}^{*}\right) \left| \nu
_{i}^{m}\right\rangle  
\nonumber \\
&\equiv &[(I+\epsilon^s)U_{MNS}^*]_{ei}\left| \nu _{i}^{m}\right\rangle,  \\
\left| \nu _{\mu }^{d}\right\rangle  &=&\sum_{i}\left( (1+\epsilon _{\mu \mu
}^{d})(U_{MNS})_{\mu i}^{*}+\epsilon _{\mu e}^{d}(U_{MNS})_{ei}^{*}
+\epsilon _{\mu
\tau }^{d}(U_{MNS})_{\tau i}^{*}\right) 
\left| \nu _{i}^{m}\right\rangle \nonumber  \\
&\equiv &[(I+\epsilon^{d})U_{MNS}^*]_{ei}\left| \nu _{i}^{m}\right\rangle. 
\end{eqnarray}
We recall that the complex conjugation comes from the fact that once
$U_{MNS}$  
is defined in the Lagrangian as given by eq.(\ref{SM}), one has \cite
{akhmedov}
\begin{equation}
\left| \nu _{e}^{f}\right\rangle =(U^{MNS})^*_{ei}\left| \nu
_{i}^{m}\right\rangle. 
\end{equation}

The observation of new physics in the detection requires not only knowledge on neutrino beams but also on quarks properties within nuclear matter \cite{Ota2001}. In particular, present uncertainties on parton distribution and hadronisation for the  range of neutrino energy  will make very difficult  to distinguish the effects of  new lepton flavour violating physics from other sources of uncertainties or new physics.  So, in the following part of the paper, we shall only compute  the effects of lepton flavour changing interactions coming from the production processes which is purely leptonic and neglect their effects in detection processes. Of course, our analysis can be trivially extended to include flavour changing effects in detection processes.

The $CP$ asymmetries in vacuum and in matter can be computed using the
standard procedure.


\subsection{Oscillations in vacuum}

It is well known that in the framework of the SM, there is no $CP$ violation
in neutrino oscillations in the limit $s_{13}\rightarrow 0$. Of course,
in the presence of physics beyond the SM, this 
is not true anymore. 
Therefore, it is interesting to study what happens in our model when $%
s_{13}\rightarrow 0$, answering the question whether even in this
limit, one can expect to observe $CP$ violation in neutrino
oscillations in models with extra isosinglet charged leptons.

It should be recalled that FCNC impose strict constraints on the 
$(U_L)_{4i}\equiv S_i$
elements of the $U_{L}$ lepton mixing matrix. 
Using the unitarity of $U_{L}$ and eq.(\ref{thetai4_def}), 
we can relate the angles $\theta_{i4}$
to the $S_i$ coefficients

\begin{equation}
\theta_{i4} e^{i\delta_{i4}} \simeq -\left( U_{\nu }^{\dagger }K\right)
_{ij}S_{j}^{\dagger }
\simeq -\left( U_{MNS}^{\dagger }\right) _{ij}S_{j}^{\dagger }.
\label{theta2}
\end{equation}
The stringent bounds on new contributions to the first family 
couplings require $S_{1}\ll S_{2,3}$. We therefore consider 
the following texture for $S$
\begin{equation}
S=(0,S_{2},S_{3}).
\end{equation}
\bigskip In such a case, $\theta _{i4}$ read
\begin{eqnarray}
\theta _{e4}e^{i\delta _{14}} &=&s_{12}\left(
c_{23}S_{2}^{*}-s_{23}S_{3}^{*}\right) +c_{12}s_{13}e^{-i\delta _{13}}\left(
s_{23}S_{2}^{*}+c_{23}S_{3}^{*}\right),   \label{phase_s13_1} \\
\theta _{\mu 4}e^{i\delta _{24}} &=&-c_{12}\left(
c_{23}S_{2}^{*}-s_{23}S_{3}^{*}\right) +s_{12}s_{13}e^{-i\delta _{13}}\left(
s_{23}S_{2}^{*}+c_{23}S_{3}^{*}\right),   \label{phase_s13_2} \\
\theta _{\tau 4} &=&-c_{13}\left( s_{23}S_{2}^{*}+c_{23}S_{3}^{*}\right). 
\label{phase_s13_3}
\end{eqnarray}
It is interesting to note from the above equations that as expected  when $s_{12} \rightarrow  0$, $\theta_{e4} \rightarrow S_1 \equiv (U_L)_{4e}$ which is strongly constrained by lepton flavour violating muon decays. But if $s_{12} \approx c_{12} \approx O(1)$, $\theta_{e4}$ is dominated by   $S_2$ and  $S_3$ contributions  which are much less constrained than $S_1$ (see eqs.(\ref{bound2})).

As discussed before, the model with one extra isosinglet charged lepton has two
additional phases. For 
simplicity, we shall choose the phases of $S_{2}$
and $S_{3}$ such that $\theta _{\tau 4}$ is real ($\delta _{34}=0$). From
these equations, the following limits can be considered 
\begin{description}
\item{(A) $s_{13}=0$} 
\begin{eqnarray}
\theta _{e4}&=&-\tan \theta_{12} \times \theta _{\mu 4}, \\
\delta _{14}&=&\delta _{24}.
\label{s13=0}
\end{eqnarray}
\item{(B) $\delta _{13}=0$ and $s_{12}=c_{12}=1/\sqrt{2}$.} We can make a
perturbative expansion in $s_{13}$ 
\begin{eqnarray}
\left| \delta _{14}-\delta _{24}\right| &\simeq& 2\frac{\theta _{\tau 4}}{%
\theta _{e4}} s_{13} \mbox{ sin}\phi +O(s_{13}^{2}), \nonumber\\
\theta_{e4,\mu4}  &\simeq& \pm \frac{a}{\sqrt{2}} \left(1 \mp s_{13} \frac{\theta_{\tau4}}{a} \mbox{ cos}\phi +O(s_{13}^{2}) \right), \label{delta13=0}\\
\mbox{sin}\delta_{14,24}  &\simeq& \mbox{sin}\phi \left(1 \pm s_{13}\frac{\theta_{\tau4}}{a} \mbox{ cos}\phi +O(s_{13}^{2})\right), \nonumber 
\end{eqnarray}
with $\phi =\mbox{ arg}(c_{23}S_{2}^{*}-s_{23}S_{3}^{*})$ and $a=\left|c_{23}S_{2}^{*}-s_{23}S_{3}^{*}\right|$. The upper and lower signs in the equation for $\theta_{e4,\mu4}$ correspond respectively to $\theta_{e4}$ and $\theta_{\mu4}$.
\end{description}

Using the relation between $\epsilon^{s}$ and $A^{NP}$ 
and expanding $A^{NP}$
in terms of $\theta _{i4}$, it is easy to get the expression for $\epsilon
^{s}$ (from now on we denote $U_{MNS}$ simply as $U$)
\begin{eqnarray}
\epsilon _{ij}^{s} &=&-U_{j1}U_{i1}^{*} \frac{\theta _{e4}^{2}}{2}%
-U_{j2}U_{i2}^{*}\frac{\theta _{\mu 4}^{2}}{2}-U_{j3}U_{i3}^{*}\frac{\theta
_{\tau 4}^{2}}{2}  
\nonumber \\
&&-U_{j2}U_{i1}^{*}\theta _{e4}\theta _{\mu 4}e^{i(\delta _{14}-\delta
_{24})}-U_{j3}U_{i1}^{*}\theta _{e4}\theta _{\tau 4}e^{i\delta
_{14}}-U_{j3}U_{i2}^{*}\theta _{\mu 4}\theta _{\tau 4}e^{i\delta _{24}}. 
\label{eps}
\end{eqnarray}

Following the formalism described in \cite{Nir2001,Ota2001},  
the dominant contribution to $CP$ asymmetries in the different
channels can be easily computed. 
We can use directly their results just replacing $\epsilon
_{ij}^{s}$ by the expressions given in eq.(\ref{eps}) and $\epsilon _{ij}^{d} \approx 0$.

Before studying the different channels of oscillations, let us
define some notation that is used later, 
\begin{eqnarray}
\Delta m_{ij}^{2} &\equiv &m_{i}^{2}-m_{j}^{2}, \\
\Delta _{ij} &\equiv &\Delta m_{ij}^{2}/(2E), \\
x_{ij} &\equiv &\Delta _{ij}L/2=\frac{\Delta m_{ij}^{2}L}{4E} \nonumber \\
&=&1.27\frac{\Delta m_{ij}^{2}}{eV^{2}}\times \frac{L}{km}\times \frac{GeV}{E%
}.
\end{eqnarray}
To get an idea of the order of magnitude of $x_{ij}$, let us evaluate them
using the data on neutrino atmospheric \cite{SuperKamiokandea,MACROa}
($\Delta m_{13}^{2}=3\times 10^{-3}$ eV$^{2}$, tan$\theta _{23}=1$) , and
for $L_{GS}=732$ km, which is the distance corresponding to the experiment
CERN-Gran Sasso\cite{Gran_Sasso}, $E=50$ GeV and in the case of LMA solution
of the solar neutrino problems (LMA parameters: $\Delta m_{12}^{2}=10^{-4}$
eV$^{2}$, tan$\theta _{12}=1$). For SMA, we used the parameters $\Delta
m_{12}^{2}=10^{-6}$ eV$^{2}$, tan$\theta _{12}=7.5\times 10^{-4}$\cite
{SuperKamiokandeb,SNOa}. 
\begin{eqnarray*}
x_{13} &\simeq &0.056, 
\\
x_{12}^{LMA} &\simeq &0.0019,
\\
x_{12}^{SMA} &\simeq &1.9\times 10^{-5}.
\end{eqnarray*}

\subsubsection{$\nu _{e}-\nu _{\mu }$ channel}

In order to distinguish the characteristic signature of this kind of models,
we first discuss the behaviour in different extreme cases
of the $CP$ asymmetries, $A_{CP}$
defined as follows 
\begin{equation}
A_{CP}\equiv \frac{P_{_{\nu _{e}-\nu _{\mu }}}-P_{_{\overline{\nu }_{e}-%
\overline{\nu }_{\mu }}}}{P_{_{\nu _{e}-\nu _{\mu }}}+P_{_{\overline{\nu }%
_{e}-\overline{\nu }_{\mu }}}}\equiv \frac{P_{e\mu }-P_{\overline{e\mu }}}{%
P_{e\mu }+P_{\overline{e\mu }}}.
\end{equation}
To get the analytical results, we should
remember that we expand $P_{e\mu }^{SM}$ to second order in $s_{13}$ and $%
P_{e\mu }^{NP}$ to first order in $s_{13}$ and we assume that $x_{21}\ll $ $%
x_{31}\ll 1\Rightarrow L\ll E_{\mu }$(GeV)$\times 262$ km. $P_{e\mu }^{SM}$
is the probability of oscillation between $\nu _{e}-\nu _{\mu }$ due to SM
and $P_{e\mu }^{NP}$ is the probability of oscillation $\nu _{e}-\nu _{\mu }$
due to New Physics.

Let us consider the case when $\delta_{13}=0$. In this case
all $CP$ violating phases are due to New
Physics. As in ref.\cite{Nir2001}, we consider 2 different limits: one
for large value of $s_{13}$ (close to the experimental bound: $\left|
U_{e3}\right| \leq 0.16$ \cite{Apollonio:1999ae}) and the other for
small $s_{13}$ (typically small means for instance, taking the LMA solution
for solar neutrinos, $s_{13}\leq 0.01$).

In the ``\underline{large}'' $s_{13}$ limit ($x_{21}/x_{31}\ll \left|
\left( U_{e3}U_{\mu 3}\right) /\left( U_{e2}U_{\mu 2}\right) \right| $), the
SM probability is given by 
\begin{equation}
P_{e\mu }^{SM}=4x_{31}^{2}\left| U_{e3}U_{\mu 3}^{*}\right| ^{2}.
\end{equation} 
The New Physics CP asymmetry
$A_{CP}^{NP}$ then reads \footnote{ As in the limit $\delta _{13}=0$, all elements of $U_{MNS}$ are real, we simplify the notation omitting the $^*$ in the following.}

\begin{eqnarray}
A_{CP}^{NP}&\simeq& -\frac{1}{x_{31}}\mbox{ Im}\left( \frac{\epsilon _{\mu
e}^{d*}+\epsilon _{e\mu }^{s}}{U_{e3}^{*}U_{\mu 3}}\right) 
\nonumber \\
%
&\simeq &\frac{1}{x_{31}}\left( \left( \frac{U_{\mu2}U_{e 1}%
}{U_{e3}U_{\mu 3}}\right) \theta _{e4}\theta _{\mu 4}\mbox{ sin}(\delta
_{14}-\delta _{24})+\theta _{\mu 4}\theta _{\tau 4}\mbox{ sin}\delta
_{24}\left( \frac{U_{e 2}}{U_{e 3}}\right) \right.   \nonumber \\
&&\left. +\theta _{e4}\theta _{\tau 4}\mbox{ sin}\delta _{14}\left( \frac{%
U_{e 1}}{U_{e 3}}\right) \right).   \label{PSM2}
\end{eqnarray}
 It is important to notice that  $A_{CP}^{NP}$ seems to be  enhanced by $U_{e3}^{-1}$ factor. In fact, using eqs(\ref{s13=0}-\ref{delta13=0}), it is easy to check that the $U_{e3}$ in denominator is cancelled once we replace the $U_{MNS}$ elements in terms of the $\theta_{ij}$ angles.   

In the ``\underline{small}'' $s_{13}$ limit ($x_{21}/x_{31}\gg \left| \left(
U_{e3}U_{\mu 3}\right) /\left( U_{e2}U_{\mu 2}\right) \right| $), the SM probability of oscillation is given by 
\begin{equation}
P_{e\mu }^{SM}=4x_{21}^{2}\left| U_{e2}U_{\mu 2}\right| ^{2}.  \label{PSM}
\end{equation}

 For the $CP$ asymmetry, we find
\begin{eqnarray}
A_{CP}^{NP} &\simeq &-\frac{1}{x_{21}}\mbox{ Im}\left( \frac{\epsilon _{\mu
e}^{d*}+\epsilon _{e\mu }^{s}}{U_{e2}^{*}U_{\mu 2}}\right)  \nonumber \\
&\simeq &\frac{1}{x_{21}}\left( \left( \frac{U_{e 1}}{U_{e 2}}%
\right) \theta _{e4}\theta _{\mu 4}\mbox{ sin}(\delta _{14}-\delta
_{24})+\theta _{\mu 4}\theta _{\tau 4}\mbox{ sin}\delta _{24}\left( \frac{%
U_{\mu 3}}{U_{\mu 2}}\right) \right.   \nonumber \\
&&\left. +\theta _{e4}\theta _{\tau 4}\mbox{ sin}\delta _{14}\left( \frac{%
U_{\mu 3}U_{e 1}}{U_{e2}U_{\mu 2}}\right) \right).   \label{PSM1}
\end{eqnarray}

In case of bimaximal mixing, eqs.(\ref{PSM2}-\ref{PSM1}) can be computed making an expansion in $s_{13}$ and using eqs(\ref{delta13=0}):
\begin{equation}
A_{CP}^{NP} \simeq \frac{1}{\sqrt{2 P_{e\mu}^{SM}}}~ a~ \theta_{\tau4} s_{13} \mbox{ sin}\phi + O(s_{13}^2). 
\label{generalACP}
\end{equation}
Both limits (``large'' and ``small'' $s_{13}$) are recovered using the appropriate $ P_{e\mu}^{SM}$.

 From this equation, it is easy to get the limit of $s_{13}= 0$. 
\begin{eqnarray}
A_{CP}^{NP}(s_{13}=0) &\sim & \frac{O(S_1S_i)}{x_{21}} \approx 0
\end{eqnarray}
with $i=2,3$.
Thus we can conclude that $CP$ asymmetries in the $\nu_{\mu}-\nu_e$ channel are unobservable for $s_{13}=0$  due to the constraint on the non-observation of rare muons flavour changing decays.

\subsubsection{$\nu _{\tau }-\nu _{\mu }$ channel}

Let us now consider $CP$ violation in the $\nu _{\tau }-\nu _{\mu }$ channel.
In this case, the interesting limit is $s_{13}\rightarrow 0$ ($\Rightarrow
A_{CP}^{SM}=0$). The SM probability is
\begin{eqnarray}
P_{\nu _{\tau }-\nu _{\mu }}^{SM} &\simeq &4\mbox{ sin}x_{31}^{2}\left|
U_{\mu 3}\right| ^{2}\left| U_{\tau 3}\right| ^{2}, 
\end{eqnarray}
leading to an induced asymmetry
\begin{eqnarray}
A_{CP}^{NP} &\simeq &\frac{1}{x_{31}}\left( \left( \frac{U_{\tau 2}U_{\mu
1}}{U_{\mu 3}U_{\tau 3}}\right) \theta _{e4}\theta _{\mu 4}%
\mbox{
sin}(\delta _{14}-\delta _{24})+\theta _{\mu 4}\theta _{\tau 4}\mbox{ sin}%
\delta _{24}\left( \frac{U_{\mu 2}}{U_{\mu 3}}\right)
\right. \nonumber \\
&&\left. +\theta _{e4}\theta _{\tau 4}\mbox{ sin}\delta _{14}\left( \frac{%
U_{\mu 1}}{U_{\mu 3}}\right) \right).
\end{eqnarray}
We see that the
last two terms are non-vanishing in the limit $\delta_{12}\sim
\delta_{24}$ and could then give observable effects at neutrino factories.
Indeed, even for $s_{13}=0$, $A_{CP}^{NP}$ can be expressed using eqs(\ref{s13=0}) as
\begin{equation}
A_{CP}^{NP}(s_{13}=0) \simeq  \frac{c_{23}}{s_{23}} a~\theta_{\tau 4} \mbox{ sin}\delta_{24}
\end{equation}

\subsection{Oscillations in matter}

In the general case, 
matter
effects~\cite{matter_effect} 
have to be taken into account in neutrino oscillations. 
They are due to forward $\nu-e$ scattering, the Lagrangian describing
them coming from the following 4-fermion interactions
\begin{equation}
\mathcal{L}_{matter}=
2\sqrt{2}G_{F}
(\delta_{\alpha e} \delta_{\beta e}+\epsilon _{\alpha \beta }^{m})
\left( \overline{\nu }%
_{\alpha }\gamma ^{\mu }P_{L}e\right) \left( \overline{e}\gamma _{\mu
}P_{L}\nu _{\beta }\right).
\end{equation}
Using the notation given above, the different coefficients read 
\begin{eqnarray}
\epsilon _{e\beta }^{m} &=&A_{e\beta }^{NP}=\epsilon _{e\beta }^{d}, \\
\epsilon _{\mu \mu ,\tau \tau }^{m} &=&\left| A_{e\mu ,e\tau }^{NP}\right|
^{2}, \\
\epsilon _{\mu \tau }^{m} &=&A_{e\mu }^{NP}A_{e\tau }^{NP*}=\epsilon _{e\tau
}^{s}\epsilon _{\mu e}^{*s}.
\end{eqnarray}
Thus we see that
the first order corrections due to New Physics appear in 
$\epsilon_{ee}^{m},\epsilon_{e\mu }^{m}$ and $\epsilon_{e\tau }^{m}$
while $\epsilon_{\mu \mu}^{m},\epsilon_{\tau \tau }^{m}$ and
$\epsilon_{\mu \tau }^{m}$ appear at higher order
in New Physics interactions.

The general expression for the probability to have $\nu_{\alpha }^{s}
\rightarrow \nu _{\beta }^{d}$ is given by 
\begin{equation}
P_{\nu _{\alpha }^{s}\rightarrow \nu _{\beta }^{d}}=\left| \left\langle \nu
_{\delta }\right| U_{\beta \delta }^{*d}e^{-iHt}U_{\alpha \gamma }^{s}\left|
\nu _{\gamma }\right\rangle \right| ^{2}.
\end{equation}
Matter effects can be described using the following Hamiltonian 
\begin{equation}
H_{\alpha \beta }=\frac{1}{2E_{\nu }}\left\{ (U_{MNS})_{\alpha i}
\left( 
\begin{array}{lll}
0 & 0 & 0 \\ 
0 & \Delta m_{21}^{2} & 0 \\ 
0 & 0 & \Delta m_{31}^{2}
\end{array}
\right) (U_{MNS})_{i\beta }^{\dagger}+
b
\left( 
\begin{array}{lll}
1+\epsilon_{ee}^{m} & \epsilon_{e\mu }^{m} & \epsilon_{e\tau }^{m} \\ 
\epsilon_{e\mu }^{m*} & \epsilon_{\mu \mu }^{m} & \epsilon_{\mu \tau }^{m} \\ 
\epsilon_{e\tau }^{m*} & \epsilon_{\mu \tau }^{m*} & \epsilon_{\tau \tau }^{m}
\end{array}
\right) \right\},
\end{equation}
with $b=2\sqrt{2}G_{F}n_{e}E_{\nu }$ and $n_{e}$ the electron density of the matter.

\section{Results}

In this section, we shall numerically estimate the values of the $CP$
asymmetries for the different cases studied in previous section, neglecting matter effects. We shall focus our analysis on the signal-to-noise ratio to emphasize on the possibility to measure such $CP$ asymmetries in neutrino oscillations, in particular for the $\nu _{e}-\nu _{\mu }$ channel, at future neutrino
factories.
For that, one proceeds as in ref.\cite{donini2000} defining the
  observable 
\begin{equation}
\overline{A}_{CP}\equiv \frac{N[\mu ^{-}]/N_{0}[e^{-}]-N[\mu
^{+}]/N_{0}[e^{+}]}{N[\mu ^{-}]/N_{0}[e^{-}]+N[\mu ^{+}]/N_{0}[e^{+}]},
\end{equation}
where 
\begin{eqnarray}
N[\mu ^{-,+}] &=&\frac{N_{\mu }N_{T}E_{\mu }}{\pi L^{2}m_{\mu }^{2}}%
\int_{E_{th}}^{E_{\mu }}12\left( \frac{E_{\nu }}{E_{\mu }}\right) ^{2}\left(
1-\frac{E_{\nu }}{E_{\mu }}\right) \sigma _{CC,\overline{CC}}(E_{\nu
})P_{e\mu }(E_{\nu })dE_{\nu },  \label{nu} \\
N_{0}[e^{-,+}] &=&\frac{N_{\mu }N_{T}E_{\mu }}{\pi L^{2}m_{\mu }^{2}}%
\int_{E_{th}}^{E_{\mu }}12\left( \frac{E_{\nu }}{E_{\mu }}\right) ^{2}\left(
1-\frac{E_{\nu }}{E_{\mu }}\right) \sigma _{CC,\overline{CC}}(E_{\nu
})dE_{\nu },  \label{ne}
\end{eqnarray}
with $\sigma _{CC,\overline{CC}}(E_{\nu })$ $=\sigma _{CC,\overline{CC}%
}E_{\nu }$ and $\sigma _{CC,\overline{CC}}$ given, respectively, by $%
0.67\times 10^{-38}$cm$^{2}/$GeV, $0.34\times 10^{-38}$cm$^{2}/$GeV. $N_{\mu
}$ is the number of useful muon decays, $N_{T}$ is the number of protons in
the target detector and $E_{th}$ the threshold energy of the detector.

The statistical error $\Delta A_{CP}$ is given by 
\begin{equation}
\Delta A_{CP}\simeq \frac{1}{\sqrt{N[\mu ^{-}]+N[\mu ^{+}]}}.
\end{equation}
The signal-to-noise ratio is given by $\overline{A}_{CP}/\Delta A_{CP}$. In
order to illustrate the possibility to detect $A_{CP}$ at future neutrino
factories, it is useful to compute the signal-to-noise ratio using the
analytical results we have for $A_{CP}$ for the different cases studied in
previous sections. These results were obtained making an expansion in $s_{13}$
(expanding $P_{e\mu }^{SM}$ to second order in $s_{13}$ and $P_{e\mu }^{NP}$
to first order in $s_{13}$) and were valid for short distances, $x_{31}\ll
1\Rightarrow L\ll E_{\mu }$(GeV)$\times 262$ km, (typically at neutrino
factories, $E_{\mu }=50$GeV \cite{Alsharoa:2002wu}). To get this number,
we use $\Delta m_{13}^{2}=3\times 10^{-3}$ eV$^{2}$.
Within these assumptions, $A_{CP}$ and $P_{e\mu }^{SM}$ have a very 
simple energy
dependence 
\begin{eqnarray}
A_{CP} &\sim &E_{\nu }, \\
P_{e\mu }^{SM} &\sim &\frac{1}{E_{\nu }^{2}}.
\end{eqnarray}
Using this energy dependence, it is easy to integrate eqs.(\ref{nu}-\ref{ne}%
). The final signal-to-noise ratio reads 
\begin{equation}
\frac{\overline{A}_{CP}}{\Delta A_{CP}}\approx A_{CP}(E_{\mu })\sqrt{P_{e\mu
}^{SM}(E_{\mu })}\left( \frac{N_{\mu }N_{T}E_{\mu }^{3}}{\pi L^{2}m_{\mu
}^{2}}\right) ^{1/2}\sqrt{\frac{\sigma _{CC}+\sigma _{\overline{CC}}}{2}}.
\end{equation}
To get an estimation, let us evaluate this expression for a total of $10^{21}$
useful muons ($N_{\mu }$) with an energy $E_{\mu }=50$ GeV and a $40$kt
detector ($N_{T}\approx 1.1\times 10^{33}$) and for $L_{GS}=732$ km. 
\begin{equation}
\frac{\overline{A}_{CP}}{\Delta A_{CP}}\approx 2\times 10^{3}A_{CP}(E_{\mu })%
\sqrt{P_{e\mu }^{SM}(E_{\mu })}\times \frac{L_{GS}}{L}.
\end{equation}

Usually, it is said that the signal could be distinguished  from the
background noise at 99\% C.L.  if  the signal-to-noise ratio is bigger than
three. Using eq.(\ref{generalACP}) and imposing   the signal-to-noise ratio to be bigger than three,  an lower bound on $a$ and $\theta_{\tau4}$ can be found,

\begin{equation}
\left( \frac{\overline{A}_{CP}}{\Delta A_{CP}}\right)>3\Rightarrow a~\theta_{\tau4} \mbox{ sin}\phi > \frac{1}{s_{13}} 2 \times 10^{-3}\frac{L}{L_{GS}}. 
\end{equation}

As $a~ \theta_{\tau4} \sim S_3^2 \approx \beta_{\tau \tau} <0.0013$ due to lepton universality (see eqs(\ref{universality})), the above lower  bound is never satisfied for future neutrino factories as they are planned \cite{Alsharoa:2002wu}.    

Thus, the constraints on FCNC in the leptonic sector and violation of lepton universality make the New Physics contribution to $CP$-asymmetries in the $\nu_e-\nu_{\mu}$ channel unobservable, in models with extra isosinglet charged leptons, with the present design of neutrino factories.


 We should stress at this point the importance of a measurement of CP
violation in the $\nu_{\mu}-\nu_{\tau}$ channel in neutrino oscillations. In this
case CP violation mediated by the new charged lepton could be
noticeable even in the limit $s_{13}=0$ what would be a clear
signature of physics beyond the SM. Indeed, proceeding in the same way than before, one gets

\begin{equation}
\left( \frac{\overline{A}_{CP}}{\Delta A_{CP}}\right)>3\Rightarrow a~\theta_{\tau4} \mbox{ sin}\delta_{24} > \frac{3}{2} \times 10^{-3}\frac{L}{L_{GS}}. 
\end{equation}

This lower bound can be translated as a lower bound on $S_3$,
\begin{equation}
S_3^2 \equiv \beta_{\tau \tau} \gtrsim \frac{3}{2} \times 10^{-3}\frac{L}{L_{GS}}.
\end{equation}
 This lower bound is still marginally compatible with the upper bound on $\beta_{\tau \tau}$ coming from lepton universality. So, in principle, there is still a small window for observation of vector-like charged leptons effects on $CP$ asymmetries in $\nu_{\mu}-\nu_{\tau}$ channel at neutrino factories. 
We should emphasise that as $B$-factories produce as many tau pairs events as $B-\bar{B}$ events, the expected improvements on rare lepton flavour changing tau decays at BELLE experiment for instance  should allow us to close this window  in a very near future if no lepton flavour changing tau decays are observed. And inversely, the observation of a such events at a $B$-factory will be a strong motivation to adapt the design of future neutrino factories to be able to probe $CP$ asymmetries in  $\nu_{\mu}-\nu_{\tau}$ channel for the kind of model studied in this paper.

\section{Conclusion}

In this paper, we have studied  models inspired by extra-dimensions where new particles 
naturally arise as isosinglet charged leptons. We have described a general perturbative approach  to compute the effects of a vector-like charged lepton on neutrino oscillations and their potential signature at future neutrino factories taking into account all constraints coming from FCNC and lepton universality. This approach is based on model-independent formalism introduced in ref. \cite{Nir2001,Ota2001} and can be applied to any extension of the SM with vector-like particles (neutrinos or charged-leptons).

We have shown that in case of leptonic maximal mixings, new physics effects in  $\nu_e-\nu_{\mu}$   $CP$ asymmetries are significantly enhanced. But due to FCNC and lepton universality constraints,  such effects  are  out of reach of current neutrino factory design. In this class of models, we can expect to observe some signal from new physics  at $1-3\sigma$  in $\nu_{\mu}-\nu_{\tau}$ channel if rare flavour changing tau decays are seen in a very near future at $B$-factories as BELLE experiment.  If such events are seen, there will be  a strong motivation to slightly adapt the neutrino factory design to improve their sensitivity to $CP$ asymmetries in   $\nu_{\mu}-\nu_{\tau}$ channel in order to test this class of models.

\section*{Acknowledgements}
It is a pleasure to thank F. del \'Aguila for usefull discussions.
This work is partially supported by the European Union 
under contract HPRN-CT-2000-00149, by MCYT under contract FPA2000-1558
and Junta de Andaluc\'\i a group FQM 101. The work of D.D.
was supported by \emph{Funda\c{c}\~{a}o para a Ci\^{e}ncia e a Tecnologia}
(FCT) through the project POCTI/36288/FIS/2000.

\appendix

\section{Kaluza-Klein expansion of bulk fermions}

As we have emphasized in the Introduction, models with extra dimensions
represent a well motivated framework for physics beyond the SM with extra
vector-like fermions. We review in this appendix the KK expansion of bulk
fermions in the class of models introduced in~\cite{DelAguila:2001pu}. Let
us consider a five-dimensional model with the fifth dimension compactified
on the orbifold $S^1/Z_2$, which is a circle of radius $R$ with the $Z_2$
identification $y\leftrightarrow -y$ or, equivalently, an interval $0\leq y
\leq \pi R$ with two boundaries, the orbifold fixed points. A fermion in
five dimensions is vector-like, the Dirac representation of $SO(1,4)$ being
irreducible, thus it admits a bare Dirac mass which, in order to be
non-trivial, has to depend on the extra dimension. %
%
%
%
%
%

The integral in the fifth dimension of the five-dimensional Lagrangian
results in the following four-dimensional Lagrangian (we use the ``mostly
minus'' convention for the metric and $\gamma^4=\mathrm{i} \gamma^5$) 
\begin{eqnarray}
\ensuremath{{\mathcal{L}}}&=&\int_0^{\pi R} \mathrm{d}y\; \bar{\Psi}\Big[ 
\mathrm{i}\gamma^N \partial_N -M(y)\Big] \Psi  \nonumber \\
&=&\int_0^{\pi R} \mathrm{d}y\; \Big[\bar{\Psi}_L \mathrm{i} \gamma^\mu
\partial_\mu \Psi_L+ \bar{\Psi}_R \mathrm{i} \gamma^\mu \partial_\mu \Psi_R 
\nonumber \\
& &\phantom{\int_0^{\pi R} \mathrm{d}y\;}+ \bar{\Psi}_R
\left(\partial_y-M(y)\right)\Psi_L+ \bar{\Psi}_L
\left(-\partial_y-M(y)\right)\Psi_R\Big],
\end{eqnarray}
where we have split the vector-like fermion into its two chiral
components $\Psi=\Psi_L+\Psi_R$ defined by $\gamma^5\Psi_{L,R}=\mp
\Psi_{L,R} $. Note that in order to have a dynamical field in the extra
dimension, that is the term $\bar{\Psi}_L \partial_y \Psi_R+\mathrm{h.c.}%
\neq 0$, the two chiralities of a fermion necessarily have opposite $Z_2$
parities to cancel the change of sign of $\partial_y$. Thus without loss of
generality we can choose the mass term to be odd $M(-y)=-M(y)$. 
In particular we take it to have a multi-kink structure
\begin{equation}
M(y)=\left\{ 
\begin{array}{l}
M,\quad 0\leq y\leq \pi a, \\ 
-M,\quad \pi a\leq y\leq \pi R,
\end{array}
\right. 
\end{equation}
with $0\leq a \leq R$. 
Hermiticity of the Lagrangian requires $M$ to be real but it can be of
either sign. To complete a four-dimensional description we expand the
five-dimensional fields in a real complete orthonormal basis 
in the fifth dimension with the coefficients of the
expansion being four-dimensional fields, the KK modes, 
\begin{equation}
\Psi_{L,R}(x,y)=\frac{1}{\sqrt{\pi R}} \sum_{n=0}^\infty f_n^{L,R}(y)
\Psi^{(n)}_{L,R}(x).
\end{equation}
Inserting this KK expansion in the Lagrangian we obtain the action of an
infinite tower of four-dimensional vector-like fermions, except for the zero
mode which will be chiral, 
\begin{eqnarray}
\ensuremath{{\mathcal{L}}}&=& \sum_n \left\{ \bar{\Psi}^{(n)}_L \mathrm{i} 
\not{\! }\partial \Psi^{(n)}_L + \bar{\Psi}^{(n)}_R \mathrm{i} \not{\!
}\partial \Psi^{(n)}_R - m_n \left[ \bar{\Psi}^{(n)}_L \Psi^{(n)}_R + \bar{%
\Psi}^{(n)}_R \Psi^{(n)}_L \right]\right\},
\end{eqnarray}
provided the following orthonormality and eigenvalue conditions are
satisfied 
\begin{eqnarray}
& &\int_0^{\pi R} \mathrm{d}y\; \frac{f_n^L f_m^L}{\pi R}= \int_0^{\pi R} 
\mathrm{d}y\; \frac{f_n^R f_m^R}{\pi R}=\delta_{n,m}, \\
& &\big[\pm \partial_y- M(y)\big]f^{L,R}_n=-m_n f_n^{R,L}.
\end{eqnarray}
The KK spectrum for this problem has been computed 
in~\cite{DelAguila:2001pu} and can be summarised as follows. (In the following
the upper (lower) sign stands for LH (RH) fields.)

There is a massless zero mode for the even chirality with exponential
localization 
\begin{equation}
f^{L,R}_0(y)=A_{L,R} \exp[\mp M |y-\pi a|],
\label{f0:fermion:flat:multibrane}
\end{equation}
where the normalization constant is given by: 
\[
A_{L,R}=\sqrt{\frac{\pm 2 M \pi R}{2-\exp[\mp 2M \pi a]-\exp[\mp 2 M \pi
(R-a)]}}. 
\]
Notice that a LH zero mode is localized at the intermediate brane for %
\mbox{$M>0$} and simultaneously localized at both orbifold fixed points with
exponential suppression in the intermediate brane (thus the designation we
use hereafter ``multi-localized'') for \mbox{$M<0$}, the opposite happens
for a RH zero mode. The rest of the spectrum is vector-like with the first
massive mode having distinct properties in the case that 
\begin{equation}
\mp 2 M \pi a(R-a)> R.  \label{multi:cond}
\end{equation}
The condition~\mbox{(\ref{multi:cond})} coincides with the
multi-localization of the zero mode (indeed we use this condition as a
quantitative definition of multi-localization in this particular problem).
In fact the multi-localization of the zero mode is intimately related to the
special properties of the first massive mode. When the condition~%
\mbox{(\ref{multi:cond})} is satisfied, the even component of the first KK
mode is also exponentially multi-localized 
\begin{equation}
f^{L,R(\mathrm{even})}_1(y)=\left\{ 
\begin{array}{l}
A\left(\ensuremath{\mathrm{e}}^{\beta_1 y}+\frac{\beta_1\mp M}{\beta_1 \pm M}
\ensuremath{\mathrm{e}}^{-\beta_1 y}\right), \quad 0\leq y \leq \pi a, \\ 
B\left(\ensuremath{\mathrm{e}}^{-\beta_1 y}+\frac{\beta_1\mp M}{\beta_1 \pm M%
} \ensuremath{\mathrm{e}}^{\beta_1 (y-2\pi R)}\right), \quad \pi a \leq y
\leq \pi R,
\end{array}
\right.
\end{equation}
where $A$ and $B$ are related by continuity of $f^\mathrm{(even)}_1$ at $%
y=\pi a$ provided the wave function does not vanish at this point and by
continuity of the derivatives if the wave function vanishes at the
intermediate brane. The parameter $\beta_1$ is the positive solution of the
eigenvalue equation (with the upper sign valid for LH components and the
lower one for RH fields) 
\begin{equation}
\beta\Big[1- \mathrm{e}^{-2\beta \pi R} \Big] = \mp M \Big[1+ \mathrm{e}%
^{-2\beta \pi R} -\mathrm{e}^{-2\beta \pi a} -\mathrm{e}^{-2\beta \pi (R-a)} %
\Big],  \label{multi:eigenvalue:eq}
\end{equation}
for even fields and the corresponding with the change $M\to -M$ for odd
fields so that the two chiralities of a fermion KK mode have the same mass
as they should. The mass of the first KK mode is $m_1^2=M^2-\beta^2_1$ and
is, provided condition~\mbox{(\ref{multi:cond})} is satisfied, always
positive and smaller than $M^2$. In particular it can be seen that in the
case of strong multi-localization the mass of the first KK modes goes
exponentially to zero 
\begin{equation}
m_1^2\approx 2 M^2 \left[ \mathrm{e}^{\pm 2\pi Ma}+\mathrm{e}^{\pm 2\pi
M(R-a)} -2\mathrm{e}^{\pm 2\pi MR}\right]\rightarrow 0\quad 
[\mp M\to \infty].
\end{equation}
It is therefore effectively decoupled from the compactification scale (thus
from the rest of the KK spectrum) in this limit. In the case that the
multi-localization condition is not satisfied this first state has the same
properties as the rest of the spectrum which consists on oscillating (thus
not localized) states with masses greater than $M$.

Let us finish this short review of the KK description of 
extra-dimensional theories writing the Yukawa couplings. The Lagrangian involving the Yukawa couplings, taking the Higgs to live at the $y=0$ boundary, reads

\begin{eqnarray}
-\lag_{Yuk} &=&
\int_0^{\pi R} \mathrm{d}y ~  \delta(y) \left\{ \lambda^{(5)} \bar{\Psi} \chi \Phi + \mathrm{h.c.} \right\} 
\nonumber \\
&=&
\frac{\lambda^{(5)}}{\pi R} \sum_{nm}
\Phi \left\{  f^{\Psi L}_n(0) f^{\chi R}_m(0) + \mathrm{h.c.} \right\},
\end{eqnarray}
 where we have chosen $\Psi_L$ and $\chi_R$ to be even fields and in the
 second equality we have expanded in KK modes.

\end{document}